\documentclass[10pt]{article}

\usepackage{amsmath}
\usepackage{array}
\usepackage{appendix}
\usepackage{graphicx}
\usepackage{amsfonts}
\usepackage{amssymb}
\usepackage{mathrsfs}
\usepackage{yfonts}
\usepackage{euscript}
\usepackage{upgreek}
\usepackage{slantsc}
\usepackage{calligra}
\usepackage[T1]{fontenc}
\usepackage{epsf}
\usepackage{latexsym}

\usepackage{tipa}

\def\be{\begin{equation}}
\def\ee{\end{equation}}
\def\beq{\begin{equation}}
\def\eeq{\end{equation}}
\def\bea{\begin{eqnarray}}
\def\eea{\end{eqnarray}}

\def\ni{\noindent}

\def\!{\hspace{-1.6667em}}

\def\mA{\mbox{A}}   
\def\mB{\mbox{B}}  
   
\def\mD{\mbox{D}}
\def\mE{\mbox{E}}

\def\mG{\mbox{G}}

\def\mJ{\mbox{J}}  

\def\mL{\mbox{L}}
\def\mM{\mbox{M}}
\def\mN{\mbox{N}}

\def\mR{\mbox{R}}

\def\mg{\mbox{g}}
\def\mh{\mbox{h}}

\def\ml{\mbox{l}}

\def\mp{\mbox{p}}

\def\fg{\mbox{\sffamily g}}

\def\fD{\mbox{\sffamily D}}

\def\scC{\mbox{\scriptsize ${\cal C}$}}          
\def\scD{\mbox{\scriptsize ${\cal D}$}}          
\def\scE{\mbox{\scriptsize ${\cal E}$}}          

\def\scG{\mbox{\scriptsize ${\cal G}$}}          
\def\scH{\mbox{\scriptsize ${\cal H}$}}          

\def\scK{\mbox{\scriptsize ${\cal K}$}}
\def\scL{\mbox{\scriptsize ${\cal L}$}}          
\def\scM{\mbox{\scriptsize ${\cal M}$}}          

\def\scP{\mbox{\scriptsize ${\cal P}$}}
\def\scS{\mbox{\scriptsize ${\cal S}$}}

\def\iB{\mbox{\scriptsize$B$}}   
\def\iD{\mbox{\scriptsize$D$}}   
\def\iK{\mbox{\scriptsize$K$}}   

\def\FrQ{\mbox{\Large $\mathfrak{q}$}}

\def\FrG{\mbox{\Large $\mathfrak{g}$}}  
 
\def\capFrG{\mbox{\boldmath$\mathfrak{G}$}}                    
\def\scapFrG{\mbox{\boldmath\scriptsize$\mathfrak{G}$}}        

\def\FA{\mbox{\Large $\mathfrak{a}$}}

\def\sg{\mbox{\scriptsize g}} 
\def\sh{\mbox{\scriptsize h}}

\def\sp{\mbox{\scriptsize p}}

\def\sE{\mbox{\scriptsize E}}

\def\sL{\mbox{\scriptsize L}}

\def\sT{\mbox{\scriptsize T}}

\def\sfg{\mbox{\sffamily{\scriptsize g}}}     

\def\sfA{\mbox{\sffamily{\scriptsize A}}}      
\def\sfB{\mbox{\sffamily{\scriptsize B}}}      
\def\sfC{\mbox{\sffamily{\scriptsize C}}}      
\def\sfD{\mbox{\sffamily{\scriptsize D}}}      
\def\sfG{\mbox{\sffamily{\scriptsize G}}}      
\def\sfK{\mbox{\sffamily{\scriptsize K}}}      
\def\sfW{\mbox{\sffamily{\scriptsize W}}}      
\def\K{Kucha\v{r} }

\def\pa{\partial}
\def\d{\textrm{d}}

\def\5Star{\mbox{\Large$\star$}}              
\def\cr{\mbox{\scriptsize{\bf $\mbox{ } \times \mbox{ }$}}}

\def\sumi2{\sum\mbox{}_{\mbox{}_{\mbox{\scriptsize $i$=1}}}^2}
\def\sumi3{\sum\mbox{}_{\mbox{}_{\mbox{\scriptsize $i$=1}}}^3}

\def\sumABcycles3{\sum\mbox{}_{\mbox{}_{\mbox{\scriptsize cycles  $A,B$=1}}}^{3}}
\def\sumCDcycles3{\sum\mbox{}_{\mbox{}_{\mbox{\scriptsize cycles  $C,D$=1}}}^{3}}

\def\sumIN{\sum\mbox{}_{\mbox{}_{\mbox{\scriptsize $I$=1}}}^{N}}

\def\sumj3{\sum\mbox{}_{\mbox{}_{\mbox{\scriptsize $j$=1}}}^3}
\def\sumk3{\sum\mbox{}_{\mbox{}_{\mbox{\scriptsize $k$=1}}}^3}

\def\Phase{\mbox{{\boldmath$\mathfrak{P}$}hase}}                     

\begin{document}

\begin{center}

{\bf \large Explicit Partial and Functional differential equations for Beables or Observables}

\vspace{.15in}

{\bf Edward Anderson} 

\vspace{.15in}

\end{center}

\begin{abstract}

We provide explicit partial differential equations -- in finite cases -- and functional differential equations  -- in field-theoretic cases -- 
which determine observables or beables in the senses of Kucha\v{r} and of Dirac.
These cover a wide range of relational mechanics models as well as Electromagnetism, Yang--Mills Theory and General Relativity.
We give an underlying reason why pure-configuration Kucha\v{r} observables are already well-known: 
various types of shape, E-fields, B-fields, loops and 3-geometries.  
The partial differential equations or functional differential equations for pure-momentum observables are also posed, 
as are those for observables which have a mixture of configuration and momentum functional dependence.

\end{abstract}

\vspace{0.1in}

\ni PACS 04.20.Cv, 04.20.Fy  

\vspace{0.1in}
  
\ni $^*$ Dr.E.Anderson.Maths.Physics *at* protonmail.com
 

\section{Introduction}

This article concerns specific physically and geometrically significant examples of constrained theories \cite{Dirac, HT92, ABook}.  
Let $Q^{\sfA}$ denote the theory's configurations, the possible values of which form its configuration space\footnote{The current article  
uses mathfrak font for spaces, so as to keep these clearly distinct from their constituent objects, and calligraphic font to pick out constraints.}
$\FrQ$.  
The corresponding conjugate momenta are denoted by $P_{\sfA}$. 
The joint space of the $Q^{\sfA}$ and $P_{\sfA}$, as equipped with the classical bracket $\mbox{\bf |[ } \mbox{ } \mbox{\bf ,} \mbox{ } \mbox{\bf ]|}$ 
constitutes phase space, $\Phase$. 

\mbox{ }

\ni Let $\scC_{\sfC} = {\cal C}_{\sfC}(Q^{\sfA}, P_{\sfA})$ 
       (or more generally $\scC_{\sfC}[Q^{\sfA,} P_{\sfA}]$: functional dependence) denote the theory's constraints.   

\mbox{ }

\ni The classical bracket is ab initio the obvious Poisson bracket, but may take a more subtle final form after due consideration of the constraints, 
such as the Dirac bracket or reduced-geometry Poisson bracket \cite{Dirac, Sni, HT92}.  

\mbox{ }

\ni{\it Beables} or {\it observables} \cite{DiracObs, Dirac, HT92, Kuchar92I93, Kuchar93, ABeables, PE-1, DO-1} are objects 
$\iB_{\sfB}$ whose `brackets' with `the constraints' $\scC_{\sfC}$ are `equal to' zero:
\be
\mbox{\bf |[}\scC_{\sfC} \mbox{\bf ,} \iB_{\sfB}\mbox{\bf ]|} \mbox{ } `='  \mbox{ } 0 \mbox{ } .
\ee
For this definition to make sense, 
a set of $\scC_{\sfC}$ such that $\mbox{\bf |[}\scC_{\sfC} \mbox{\bf ,} \, \scC_{\sfC^{\prime}}\mbox{\bf ]|}$ closes is required \cite{ABeables}.    
Beables or observables\footnote{I also use an extension from the notion of observables, 
which eventually carry nontrivial connotations of `are observed', to {\it beables}, which just `are'.
The latter are somewhat more general, so as to cover a number of viable `realist' approaches at the quantum level \cite{ABeables}.
Note that this generalization concerns not a change of definition but rather a more inclusive context in which the entities are interpreted.} 
are more useful than just any functionals of the $Q^{\sfA}$'s and $P_{\sfA}$'s through solely containing physical information.  
This property is, at the very least, required in phrasing final answers to physical questions about a theory.

\mbox{ }

\ni There are moreover a number of different possibilities for which constraints, 
which brackets and even which notion of equality can be involved in the definition.    
Usually Dirac's notion \cite{Dirac} of weak equality $\approx$ is assumed. 
Dirac \cite{DiracObs} and \K \cite{Kuchar93} notions of beables or observables apply to the range of theories considered here.  
These involve commuting with, respectively, all first-class constraints and all first-class linear constraints.
Note that Dirac = \K for Particle Physics' gauge theories, since in this context linear constraints are all.  
For relational particle mechanics (RPMs) and General Relativity (GR), however, there is a quadratic constraint too, so the two notions are not the same.
On the other hand, Supergravity theories require dropping or replacing the latter notion \cite{ABeables, AGates, AMech}, 
due to their quadratic constraint now being an integrability of their supersymmetric constraint.

\mbox{ } 

\ni I reviewed beables and observables in \cite{ABeables} (see e.g. \cite{Kuchar92I93, Kuchar93, DittrichThiemannBookTambornino, APoT3} 
for other reviews partly or totally on this topic, including the resulting `Problem of Beables' facet of the Problem of Time \cite{Kuchar92I93, ABook}).
The current article lays out intermediate step: explicit partial or functional DEs (PDEs and FDEs) for classical beables and observables.  
Here the bracket involved is specifically a Poisson bracket  
(or some generalization thereof, such as the Dirac bracket \cite{Dirac, HT92} or the Poisson bracket corresponding to an extended phase space \cite{BT91, HT92}).  
In the finite case, beables obey the PDE 
\be
\sum_{\sfA} 
\left\{
\frac{\pa\scC_{\sfC}}{\pa Q^{\sfA}}\frac{\pa \iB_{\sfB}}{\pa P_{\sfA}} - \frac{\pa\scC_{\sfC}}{\pa P_{\sfA}}\frac{\pa \iB_{\sfB}}{\pa Q^{\sfA}}
\right\}  \mbox{ } `='  \mbox{ } 0 \mbox{ } .
\ee
In the field-theoretic case, one has instead the FDE 
\be
\int \d^n z \, \sum_{\sfA}
\left\{
\frac{\delta(\scC_{\sfC}|\pa \xi^{\sfC})}{\delta Q^{\sfA}(z)}  \frac{\delta (\iB_{\sfB}|\chi^{\sfB})}{\delta P_{\sfA}(z)} - 
\frac{\delta(\scC_{\sfC}|\delta \xi^{\sfC})}{\delta P_{\sfA}(z)}  \frac{\delta (\iB_{\sfB}|\chi^{\sfB})}{\delta Q^{\sfA}(z)}
\right\}  \mbox{ } `='  \mbox{ } 0  \mbox{ } .
\label{2}
\ee
This makes use of {\it smearing}, cast in an inner product type form \cite{AM13} ( | ):  
\be 
(    \scC_{\sfW}    |    \mA^{\sfW}    )    :=    \int d^3z \, \scC_{\sfW}(z^i; Q^{\sfA}(z^j), P_{\sfA}(z^j)] \, \mA^{\sfW}(z^i) \mbox{ } . 
\ee
Beables equations are, fortunately, quite simple FDEs in some relevant senses; see Appendix A for a few simple results about this.   

\mbox{ }

\ni I next present PDEs and FDEs for pure configuration \K beables or observables as well as for pure momentum ones. 
Solutions to the former are already known, whereas solutions to the latter remain largely a mathematical problem posed in the present article, 
as is the matter of Dirac beables.
More specifically, the current article considers (Sec 2) new examples of \K beables from the new extended set of RPMs presented in \cite{AMech, ASphe} 
(in addition to the earlier RPMs \cite{BB82, B03} whose beables were already considered in \cite{ABeables} and in more detail for 3 and 4 particles in 2-$d$ in \cite{FileR, QuadI}).  
Sec 3 recollects beables or observables in Electromagnetism and Yang--Mills Theory for useful comparison.
Finally, Sec 4 provides the explicit equations for Dirac and \K beables or observables for GR, 
including posing the pure momentum beables FDE for this case of \K beables or observables.

\section{Relational particle model beables or observables}

\ni Our objective is to consider the theories in Fig \ref{More-G-Lie-2} corresponding to the geometrically significant groups in Fig \ref{More-G-Theories}.
The corresponding configuration beables are tabulated in Fig \ref{More-G-Invariants}, with the corresponding configuration spaces listed in Fig \ref{More-G}.  
%
{            \begin{figure}[ht]
\centering
\includegraphics[width=1.0\textwidth]{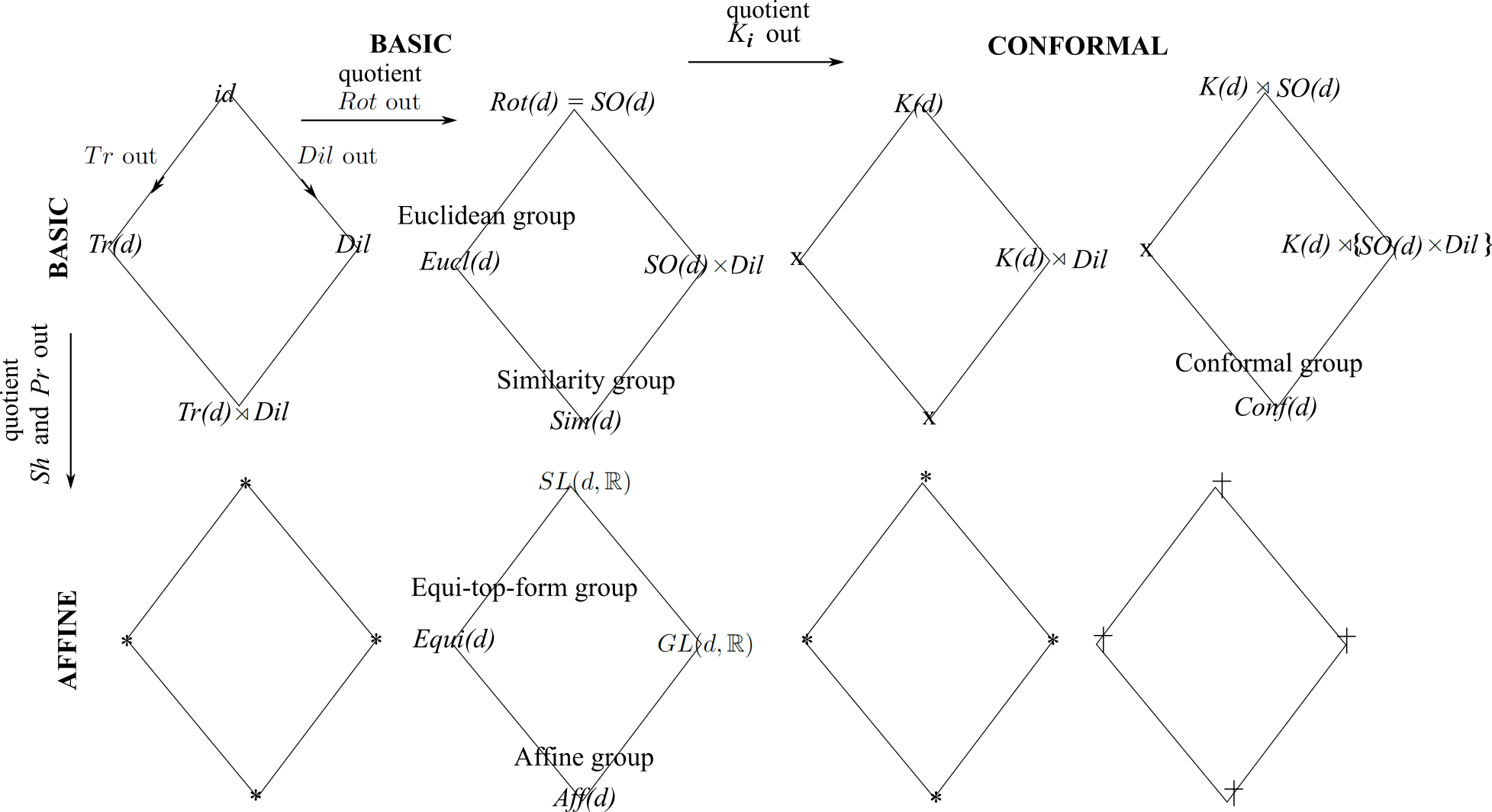}
\caption[Text der im Bilderverzeichnis auftaucht]{        \footnotesize{Summary sketch, including further groups acting upon $\mathbb{R}^d$ by quotienting.  
$Tr$ are translations, $Rot$ are rotations, and $Dil$ are dilations. 
I use $P$, $M$ and $D$ for their generators respectively, alongside $K$ for the special conformal transformations', $Sh$ for the shears and  $Pr$ for the `Procrustean stretches' 
(The last is a top form preserving stretch, for the top form supported by the dimension in question, e.g. area-preserving in 2-$d$ or volume-preserving in 3-$d$.)
Using the abstract Lie group form of brackets, then absences marked $X$       are due to the integrability $\mbox{\bf |[} K  \mbox{\bf ,} \,  P \mbox{\bf ]|} \sim D + M$. 
Absences marked $*$       are due to the integrability $\mbox{\bf |[} Sh \mbox{\bf ,} \, Sh \mbox{\bf ]|} \sim M$.
Finally, absences marked $\dagger$ are due to affine-to-special-conformal obstruction.} }
\label{More-G-Lie-2} \end{figure}          }
%
{            \begin{figure}[ht]
\centering
\includegraphics[width=1.0\textwidth]{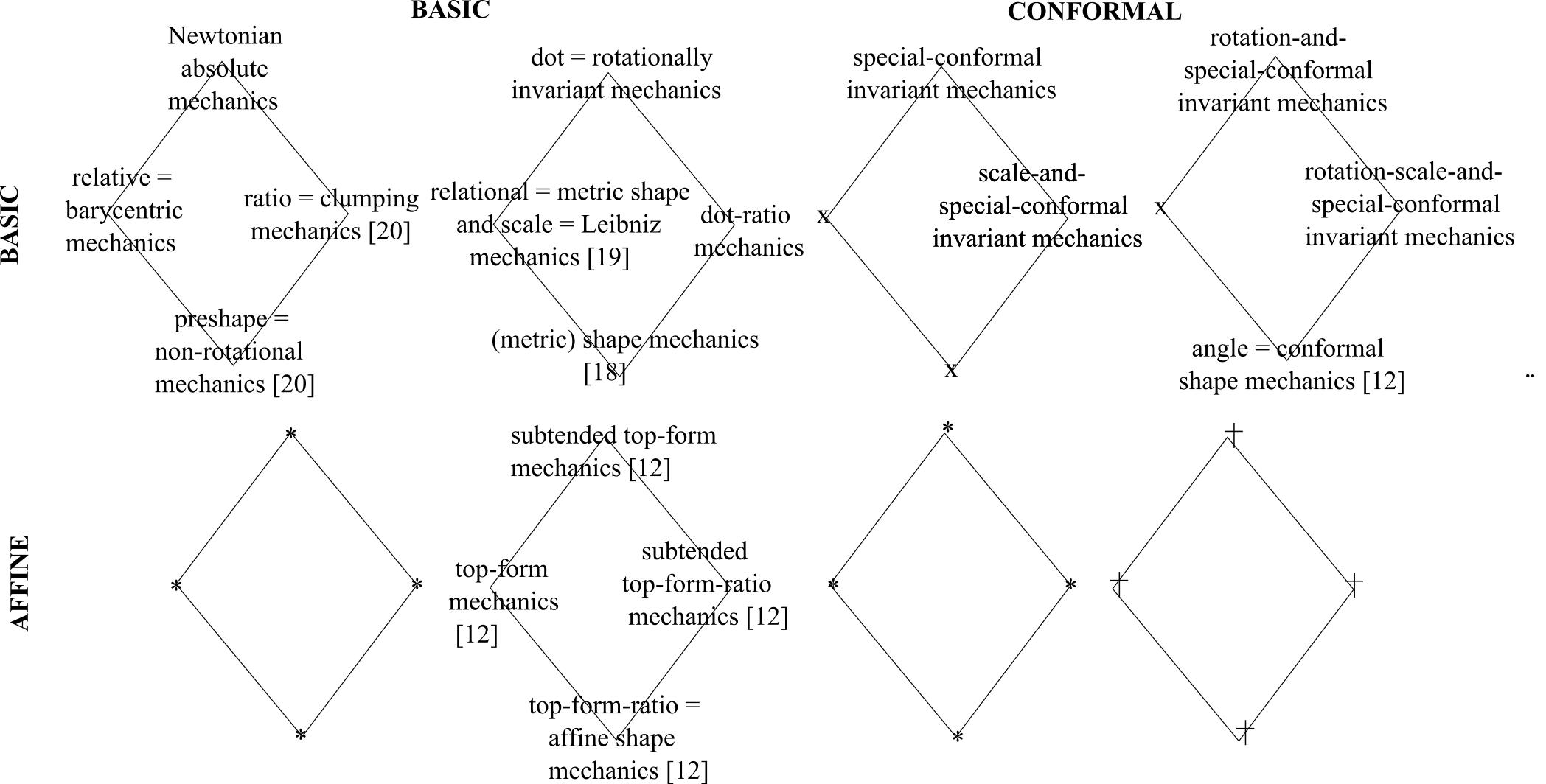}
\caption[Text der im Bilderverzeichnis auftaucht]{        \footnotesize{Corresponding RPM theories' names.  
%
} }
\label{More-G-Theories} \end{figure}          }

\ni 1) The {\it zero total momentum constraint} is 
\be
\underline{\scP} := \sumIN \underline{q}^I = 0    \mbox{ } .   
\ee
The corresponding classical \K beables condition $\mbox{\bf \{} \underline{\scP} \mbox{\bf ,} \, \iK_{\sfK} \mbox{\bf \}} \mbox{ } `='  \mbox{ } 0$ gives the PDE
\beq
\sumIN \frac{\pa \iK_{\sfK}}{\pa \underline{q}^I} \mbox{ } `='  \mbox{ } 0 \mbox{ } ,
\label{K-P}
\eeq 
which is solved by the relative interparticle separation vectors and linear combinations thereof. 
Indeed the latter are often more convenient, 
in particular the well-known {\it relative Jacobi coordinates} $\rho^A$ \cite{Marchal} whose interpretation is as relative interparticle cluster separation vectors. 

\mbox{ }

\ni 2) The {\it zero total angular momentum constraint} is 
\be
\underline{\scL} := \sumIN \underline{q}^I \cr \underline{p}_I = 0
\ee 
(or the 3-component of this in 2-$d$). 
The corresponding \K beables condition 
\be
\mbox{\bf \{} \scL \mbox{\bf ,} \, \iK_{\sfK} \mbox{\bf \}} \mbox{ } `='  \mbox{ } 0
\ee 
then gives the PDE
\beq
\sumIN 
\left\{ 
\frac{\pa \iK_{\sfK}}{\pa \underline{p}^I} \cr \underline{p_I} \mbox{ } \mbox{ } + \mbox{ } \mbox{ } \underline{q}^I \cr \frac{\pa \iK_{\sfK}}{\pa \underline{q}^I}
\right\} \mbox{ } `='  \mbox{ } 0 \mbox{ } .  
\label{K-L}
\eeq
This is solved by various dot products. 
In particular the pure-configurational \K beables equation 
\beq
\sumIN \underline{q}^I \cr \frac{\pa \iK_{\sfK}}{\pa \underline{q}^I} \mbox{ } `='  \mbox{ } 0
\eeq
is solved by 
\be
\underline{q}^I \cdot \underline{q}^J \mbox{ } ,
\ee 
and the pure-momentum \K beables equation
\beq
\sumIN \frac{\pa \iK_{\sfK}}{\pa \underline{p}_I} \cr \underline{p}_I \mbox{ } `='  \mbox{ } 0
\eeq
is solved by 
\be 
\underline{p}^I \cdot \underline{p}^J
\ee 
due to $\underline{q}^I \leftrightarrow \underline{p}^I$ exchange symmetry with the previous equation.
The full equation is {\sl not} solved by $\underline{q}^I \cdot \underline{p}^J$ but by 
\be
\underline{q}^I \cdot \underline{p}^J + \underline{p}^I \cdot \underline{q}^J \mbox{ } ,
\ee 
which is of course the outcome of applying the product rule to $\underline{q}^I \cdot \underline{q}^J$. 
Note that norms and angles are particular cases among the above, once the Appendix's Lemma 1 is taken into account.

\mbox{ }

\ni The Appendix's Remark 2 moreover also holds: if (\ref{K-P}) and (\ref{K-L}) both apply, then the solutions are dots of differences, 
which is often moreover usefully rephrased as dots of relative Jacobi vectors.
These are the \K beables \cite{ABeables} for metric shape-and-scale RPM \cite{BB82}.  

\mbox{ }

\ni 3) The {\it zero total dilational momentum constraint} is 
\be
\scD := \sumIN \underline{q}^I \cdot \underline{p}_I = 0                  \mbox{ } .
\ee
The corresponding \K beables condition 
\be 
\mbox{\bf \{} \scD \mbox{\bf ,} \, \iK_{\sfK} \mbox{\bf \}}\mbox{ } `='   \mbox{ } 0
\ee 
then gives the PDE
\beq
\sumIN 
\left\{ 
\frac{\pa \iK_{\sfK}}{\pa \underline{p}^I} \cdot \underline{p_I} \mbox{ } \mbox{ } + \mbox{ } \mbox{ } \underline{q}^I \cdot \frac{\pa \iK_{\sfK}}{\pa \underline{q}^I}
\right\} \mbox{ } `='  \mbox{ } 0                                         \mbox{ } . 
\label{11}
\eeq
The above can be recognized as an Euler's homogeneity equation of degree zero, so its solutions are ratios.  
The pure-configurational \K beables equation is then
\beq
\sumIN \underline{q}^I \cdot \frac{\pa \iK_{\sfK}}{\pa \underline{q}^I} \mbox{ } `='  \mbox{ } 0 \mbox{ } , 
\eeq
and the pure-momentum \K beables equation is (note the $\underline{q}^I \leftrightarrow \underline{p}^I$ exchange symmetry again)
\beq
\sumIN \frac{\pa \iK_{\sfK}}{\pa \underline{p}_I} \cdot \underline{p}_I \mbox{ } `='  \mbox{ } 0 \mbox{ } . 
\label{K-D}
\eeq
\ni The Appendix's composition principle then gives that if (\ref{K-P}) and (\ref{K-D}) apply, have ratios of differences, 
if (\ref{K-L}) and (\ref{K-D}) apply, have ratios of dots, and if all three apply, have ratios of dots of differences. 
These are the \K beables \cite{ABeables} for metric shape RPM \cite{B03}.  
See \cite{Kendall, FileR, QuadI} for more on metric pure shapes; \cite{AConfig} contains a summary of the corresponding configuration spaces for these, 
with comparison to those of GR.

\mbox{ } 

\ni 4) One further possibility involves a $\mathbb{R}^d \longrightarrow \mathbb{S}^d$ change of underlying absolute space model.
In the case modulo isometries, the constraints are 
   $\scL$                                         for $\mathbb{S}^1$, 
   $\underline{\scL}$                              in $\mathbb{S}^2$,  
and a pair $\underline{\scL}$, $\underline{\scL}^{\prime}$ in $\mathbb{S}^3$. 
The unreduced variables are now $\mathbb{S}^p$ angles $\theta^{iI}$ rather than particle positions $q^{iI}$ are involved, and there are no $Tr(d)$ or $Dil$.
The invariants in this case are the spherical version of the dot product.

\mbox{ }

\ni 5) The RPM on the torus $\mathbb{T}^n$ modulo isometries has also been worked out \cite{Shapes-I, ATorus}. 

\mbox{ } 

\ni 6) A second branch of further consistent possibilities \cite{AMech} involves extending the zero total angular momentum constraint 
to the {\it zero total} $SL(d, \mathbb{R})$ {\it momentum constraint} 
\be
\underline{\scS} := \sumIN \underline{q}^I \underline{\underline{\underline{S}}} \mbox{ } \underline{p}_I \mbox{ } .
\ee 
E.g. in 2-$d$, 
\beq
\underline{\underline{\underline{S}}} := \mbox{\Huge[} \mbox{\Huge(}\stackrel{\mbox{1 \,\, 0}}{\mbox{0 \, --1}}      \mbox{\Huge)},
                     \mbox{\Huge(}\stackrel{\mbox{0 \,\, 1}}{\mbox{1 \,\,\, 0}}   \mbox{\Huge)}, 
	                 \mbox{\Huge(}\stackrel{\mbox{0 \, --1}}{\mbox{1 \,\,\, 0}}    \mbox{\Huge)}\mbox{\Huge]}^{\sT} \mbox{ }  ;
\label{SL-Array}
\eeq
(20) is an area-preserving constraint in this case.
In 3-$d$, it is a volume-preserving constraint, and in general it is a top form preserving constraint for the top form corresponding to the dimension in question.  
Then the classical \K beables condition 
\be 
\mbox{\bf \{} \scS \mbox{\bf ,} \, \iK_{\sfK} \mbox{\bf \}} \mbox{ } `='  \mbox{ } 0
\ee 
gives the PDE    
\beq
\sumIN 
\left\{ 
\frac{\pa \iK_{\sfK}}{\pa \underline{p}^I} \mbox{ } \underline{\underline{\underline{S}}} \mbox{ } \underline{p_I}                            \mbox{ } \mbox{ } - \mbox{ } \mbox{ }  
                           \underline{q}^I \mbox{ } \underline{\underline{\underline{S}}} \mbox{ } \frac{\pa \iK_{\sfK}}{\pa \underline{q}^I}
\right\} \mbox{ } `='  \mbox{ } 0 \mbox{ } .
\eeq
This is solved in 2-$d$ by areas between pairs vectors, 
in 3-$d$ by volumes of parallelepipeds formed by triples of vectors, 
and in arbitrary $d$ by the top form supported by that dimension formed by $d$-tuplets of such vectors.  
The pure-configuration \K beables equation is 
\beq
\sumIN 
\underline{q}^I \mbox{ }  \underline{\underline{\underline{S}}}  \mbox{ }  \frac{\pa \iK_{\sfK}}{\pa \underline{q}^I} \mbox{ } `='  \mbox{ } 0 \mbox{ } ,
\eeq
and the pure-momentum \K beables equation is 
\beq
\sumIN 
\frac{\pa \iK_{\sfK}}{\pa \underline{p}^I}  \mbox{ }  \underline{\underline{\underline{S}}}  \mbox{ }  \underline{p_I} \mbox{ } `='  \mbox{ } 0 \mbox{ } 
\eeq
(note $\underline{q}^I \leftrightarrow \underline{p}^I$ exchange symmetry once again). 
The Appendix's composition principle continues to apply here, so we can have top forms of differences, ratios of top forms and ratios of top forms of differences. 
The last of these corresponds to affine geometry and the first of these to `equi-top-form' geometry (equiareal \cite{Coxeter} in 2-$d$).  

\mbox{ }

\ni 7) A third branch involves including instead the special conformal transformations
This follows from the 19th century observation that inversion in sphere also preserves angles, 
and well-known in 20th century Theoretical Physics through e.g. Conformal Field Theory (CFT) and its subsequent Particle Physics and String Theory applications.
Moreover, this is no longer compatible with trivial removal of translations. 
In the RPM context, including the special conformal transformations leads to the {\it zero total special conformal constraint} \cite{AMech}
\be
\scK_a := \sumIN \{q^{I\,2} {\delta_a}^b - 2q^I_a q^{Ib}\}p_{Ib} = 0 \mbox{ } . 
\ee
Additionally, four subgroups including special conformal transformations are indicated in Fig \ref{More-G-Lie-2}. 
All require the invariants to be angles (special conformal transformation is a strong condition in this regard). 
The case with $Tr(d)$ as well does use the $\rho^{iA}$ version; the other cases jointly involve the $q^{iI}$ version.  

\mbox{ } 
  
\ni The classical \K beables equation 
\be 
\mbox{\bf \{} \scK_a \mbox{\bf ,} \, \iK_{\sfK} \mbox{\bf \}} \mbox{ } `='  \mbox{ } 0
\ee 
then gives  
\beq
\sumIN
\left\{ 
2\{ 2 \, p_{I[i}q_{j]I} - (\underline{q}\cdot\underline{p}) \delta_{ij}\}  \frac{\pa \iK_{\sfK}}{\pa p_{Ij}} - 
\{q_I^2 \delta^{ij} - 2q^{iI}q^{Ij}\}                                 \frac{\pa \iK_{\sfK}}{\pa q^{Ij}}
\right\} \mbox{ } `='  \mbox{ } 0 \mbox{ } .
\eeq
The pure-configuration \K beables equation is 
\beq
\sumIN \{q_I^2 \delta^{ij} - 2 \, q^{iI}q^{Ij}\}                                 \frac{\pa \iK_{\sfK}}{\pa q^{Ij}} \mbox{ } `='  \mbox{ } 0 \mbox{ } . 
\eeq
The pure-momentum \K beables equation is 
\beq
\sumIN 2\{2 \, p_{I[i}q_{j]I} - (\underline{q}^I\cdot\underline{p}_I) \delta_{ij}\}  \frac{\pa \iK_{\sfK}}{\pa p_{Ij}} \mbox{ } `='  \mbox{ } 0 \mbox{ } 
\eeq
(note that this is no longer symmetric with the previous equation).  

{            \begin{figure}[ht]
\centering
\includegraphics[width=0.9\textwidth]{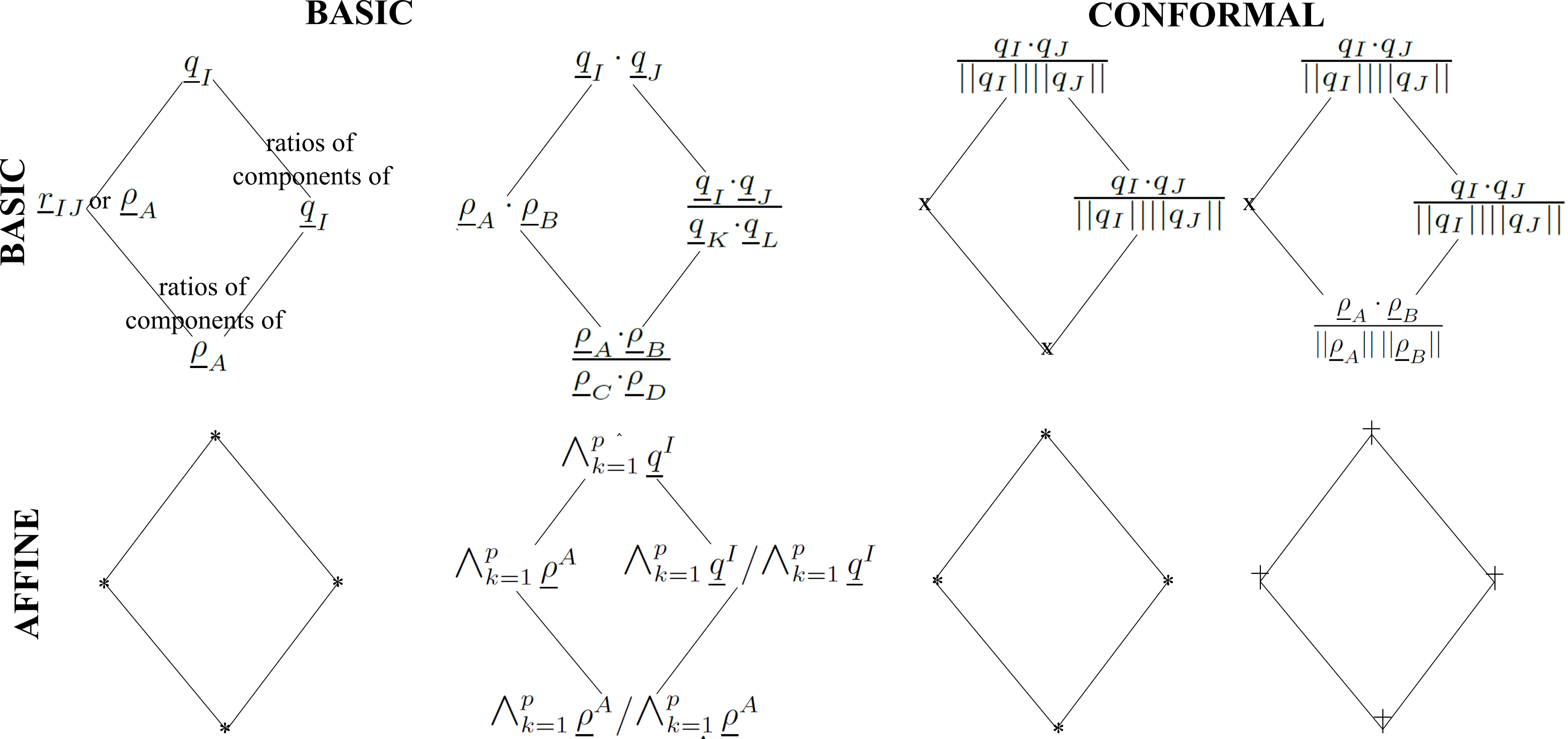}
\caption[Text der im Bilderverzeichnis auftaucht]{        \footnotesize{ \ni $\FrG$-invariants. } }
\label{More-G-Invariants} \end{figure}          }
%
{            \begin{figure}[ht]
\centering
\includegraphics[width=1.0\textwidth]{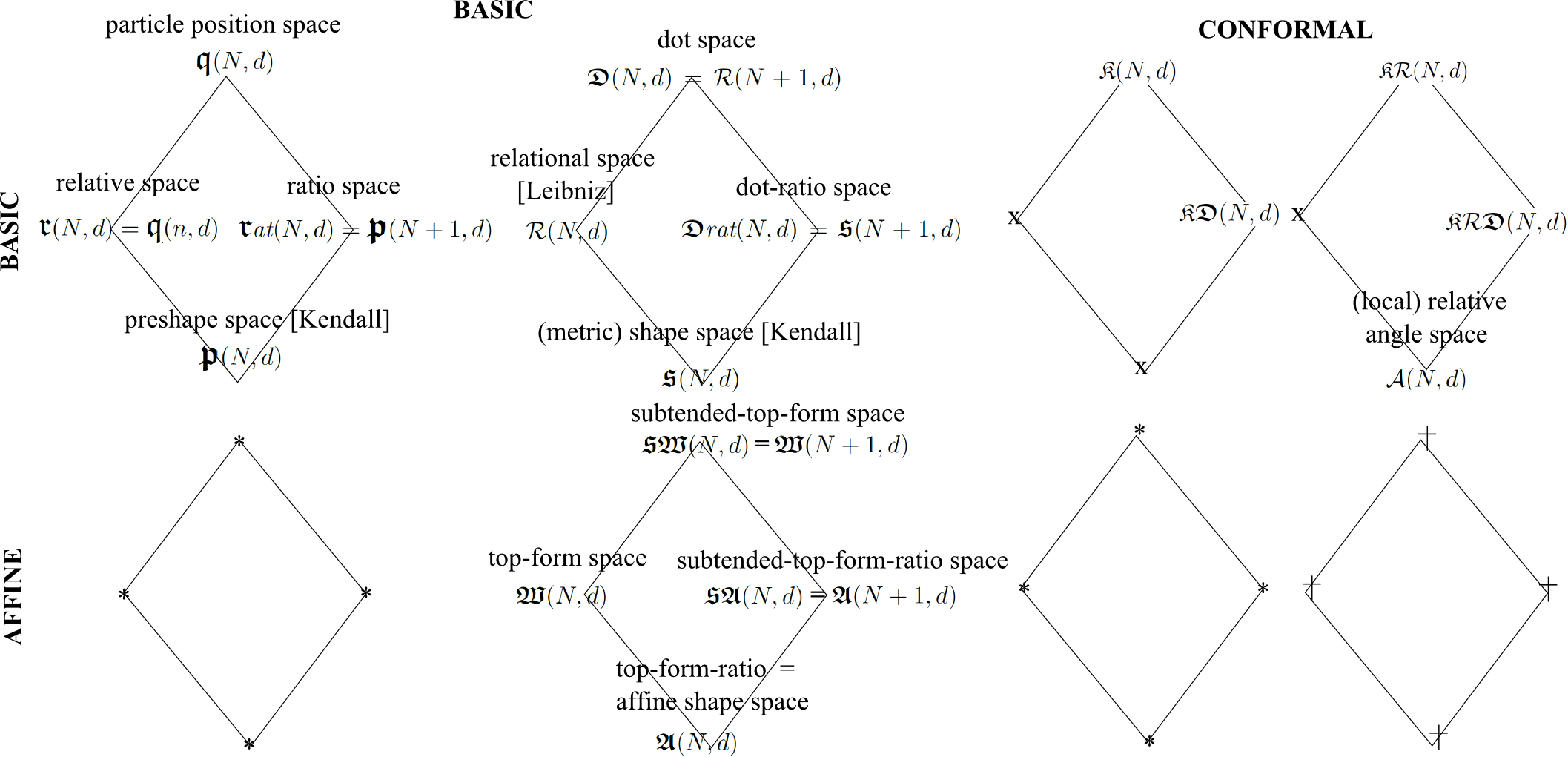}
\caption[Text der im Bilderverzeichnis auftaucht]{        \footnotesize{ \ni  Corresponding relational configuration spaces \cite{AMech}. 
Useful inter-relations are also laid out: a number of configuration spaces are the same as others with one particle less by $n$ Jacobi coordinates 
taking the same mathematical form as $N$ point particle coordinates. 
`Leibniz' here refers to the relational side \cite{L} of the venerable Absolute versus Relational (Motion) Debate, whereas `Kendall' refers to Kendall's Shape Theory \cite{Kendall}.} }
\label{More-G} \end{figure}          }

\ni Remark 1) In 2-$d$, the full conformal group becomes infinite-dimensional and thus unsuitable for finite RPMs. 

\mbox{ } 

\ni Remark 2) One can also consider the sphere versions of stripping away further layers of geometrical structure, though I leave this matter for a future occasion.  

\mbox{ } 

\ni Dirac beables equations for the theories involving whichever subgroup-forming combination of $\scP$, $\scL$, $\scD$ and $\scK$ the following extra equation.  
\be
0 \mbox{ } `='  \mbox{ } \mbox{\bf \{}\scE \mbox{\bf  ,} \, \iD_{\sfD}  \mbox{\bf \}} 
            =   \sumIN
\left\{
\frac{\pa V}{\pa q^I}\frac{\pa \iD_{\sfD}}{\pa p_I} - p_I \frac{\pa \iD_{\sfD}}{\pa q^I}
\right\}                                                                                          \mbox{ } .  
\ee
A distinct equation is required in the other cases. 
E.g.\ the equiareal and 2-$d$ affine cases each require distinct $\scE$'s built from cross-products rather than from Euclidean norms.

\section{Gauge Theory beables or observables}

For Electromagnetism, which has the {\it Gauss constraint} 
\be
\sfG := \underline{\pa} \cdot \underline{\pi}                                                      \mbox{ } ,
\ee 
the classical \K beables condition 
\be
\mbox{\bf \{} (\scG| \xi) \mbox{\bf ,} \, (\iK_{\sfK}|\chi^{\sfK}) \mbox{\bf \}} \mbox{ } `='  \mbox{ }  0
\ee 
(for smearing functions $\xi$ and $\chi^{\sfK}$) gives the FDE
\beq
\underline{\pa}\cdot\frac{\delta \iK_{\sfK}}{\delta \underline{\mA}} \mbox{ } `='  \mbox{ }  0 \mbox{ } .
\eeq
This is solved by $\underline{\mE}$ and $\underline{\mB} = \underline{\pa} \cr \underline{\mA}$, and thus by a functional 
\be 
{\cal F}[\underline{\mB}, \, \underline{\mE}]
\ee 
by Lemma 1 in the Appendix.
We can also write this in an integrated version in terms of fluxes: 
\be
{\cal F}
\left[
\iint_S\underline{\mB}\cdot \d\underline{S}, \iint_S\underline{\mE}\cdot \d\underline{S}
\right] 
= {\cal F}[W_{\gamma}, \Phi_S^{\sE}]
\ee
for electric flux $\Phi_S^{\sE}$ and loop variable 
\be 
W_{\gamma} := \mbox{exp}
\left(
\oint_{\gamma} \underline{\mA}\cdot \d\underline{\ml}
\right) 
\mbox{ } .  
\ee
This is by use of Stokes' Theorem with $\gamma := \pa S$ and then insertion of the exponentiation function subcase of Lemma 1 
(this ties the construct to the geometrical notion of holonomy).

\mbox{ } 

\ni All of the above carries over to Yang--Mills theory as well. 
This has the {\it Yang--Mills--Gauss constraint}\footnote{Here $\sfg$ is the coupling constant, $f_{IJK}$ are structure constants, $\sg^P$ are group generators, 
$\fD$ the corresponding fibre bundle notion of covariant derivative, and $P$ is the path-ordering symbol.}
\be
{\scG}_J := \fD_a\uppi^a_J = \pa_a\uppi^a_J - \mbox{\fg} f_{IJK}\mA^K_a\pi^{Ia} = 0 \mbox{ } .  
\ee
The classical \K beables condition 
\be 
\mbox{\bf \{} (\scG_I| \xi^I) \mbox{\bf ,} \, (\iK_{\sfK}|\chi^{\sfK}) \mbox{\bf \}} \approx 0
\ee 
gives 
\beq
\underline{\fD_a}\cdot\frac{\delta \iK_{\sfK}}{\delta \mA_{aJ}} \approx 0 \mbox{ } ,  
\eeq
which is solved by $\underline{\mE}_I$ and $\underline{\mB}_I$, so 
\be
{\cal F}[\underline{\mE}_I, \, \underline{\mB}_I]
\ee
solves as well by Lemma 1.  
Once again, this can be rewritten as 
\be 
{\cal F}[W_{\gamma}, \Phi_S^{\sE}] \mbox{ } ,
\ee 
now for loop variable \cite{GPBook}
\be 
W(\gamma) := \mbox{Tr}\left(P \, \mbox{exp}\left( i \fg \oint_{\gamma} \d x^i \mA_{iP}(x) \mg^P(x) \right)\right) \mbox{ } .  
\ee

\section{GR beables or observables}

For GR-as-geometrodynamics \cite{ADM},\footnote{$\mh_{ij}$ has determinant $\sqrt{\mh}$, covariant derivative $\mD_i$, Ricci tensor $\mR_{ij}$, Ricci scalar $\mR$, 
and conjugate momentum $\mp^{ij}$ with trace $\mp$. 
$\Lambda$ is the cosmological constant.} 
the linear {\it GR momentum constraint} is 
\beq
\scM_i = - 2 \, \mD_j {\mp^j}_i = 0 \mbox{ } .
\eeq 
The corresponding \K beables condition is then\footnote{Here $\pounds_{\sL}$ denotes the Lie derivative with respect to $\sL_i$.} 
\beq
\mbox{\bf \{} (    \scM_i    |    \mL^i    )   \mbox{\bf ,} \, (    \iK_{\sfK}    |    \chi^{\sfK}    )   \mbox{\bf \}}  = 
\left(  
\left\{  
\pounds_{\sL} \mh_{ij} \, \frac{\delta}{\delta \mh_{ij}} + 
\pounds_{\sL} \mp^{ij} \, \frac{\delta}{\delta \mp^{ij}} 
\right\} 
\iK_{\sfK} \, \mbox{\Huge |} \, \chi^{\sfK} 
\right) 
\mbox{ } `='  \mbox{ } 0 \mbox{ } , 
\label{Gdyn-K}
\eeq
which corresponds to the unsmeared FDE 
\beq
2 \, \mh_{jk}  \mD_i  \frac{\delta\iK_{\sfK}}{\delta \mh_{ij}}                                                                                             + 
\big\{      \mD_i  \mp^{lj} - 2{\delta^{j}}_i  \{  \mD_e \mp^{le}  +  \mp^{le}\mD_e  \}    \big\} \, \frac{\delta\iK_{\sfK}}{\delta \mp^{lj}} \mbox{ } `='  \mbox{ } 0 \mbox{ } . 
\label{GR-KB}
\eeq
In the weak case, we can furthermore discard the penultimate term. 
Then the purely configurational solutions of 
\be
2 \,   \mh_{jk}  \mD_i\frac{\delta\iK_{\sfK}}{\delta \mh_{ij}}  \mbox{ } `='  \mbox{ } 0 \mbox{ } 
\label{GR-KB-Q}
\ee
are 3-geometry quantities `${\capFrG}^{(3)}$' by (\ref{GR-KB-Q}) emulating (and moreover logically preceding) the quantum momentum constraint, 
\be
2 \, \mh_{jk}  \mD_i  \frac{\delta\Psi}{\delta \mh_{ij}} = 0 
\ee
\cite{Battelle}. 
Moreover, explicit `basis beables' (see the Appendix) are not known in this case.  

\mbox{ } 

\ni On the other hand, the complementary part of this gives a FDE for the associated 3-geometry momenta `$\Pi^{\scapFrG^{(3)}}$'.  
These formal entities solve the the GR momentum beables equation 
\ni\beq
\big\{    \mD_i \mp^{lj} - 2 \, {\delta^{j}}_i  \{  \mD_e \mp^{le} + \mp^{le}\mD_e  \}    \big\} \, \frac{\delta\iK_{\sfK}}{\delta \mp^{lj}} \mbox{ } `='  \mbox{ } 0 \mbox{ } . 
\label{GR-KB-P}
\eeq
\ni N.B. that GR also has a quadratic {\it Hamiltonian constraint} 
\be
\scH := \mN_{ijkl}\mp^{ij}\mp^{kl} - \sqrt{\mh}\{  \mR - 2\Lambda\}
\ee
Here 
\be 
\mM^{abcd} := \sqrt{\mh}\{\mh^{ac}\mh^{bd} - \mh^{ab}\mh^{cd}\}
\ee              
is the GR kinetic metric, and       
\be 
\mN_{abcd} := \{\mh_{ac}\mh_{bd} - \mh_{ab}\mh_{cd}/2\}/\sqrt{\mh}
\ee 
is its inverse the DeWitt supermetric \cite{DeWitt67}.
Moreover, under the DeWitt 2-index to 1-index map \cite{DeWitt67}, 
\be
\mh_{ab} \mbox{ } \longrightarrow \mbox{ } \mh^A \mbox{ } \mbox{ and } \mbox{ } \mp^{ab} \mbox{ } \longrightarrow \mbox{ } \mp_A \mbox{ } ,
\ee
\be   
\mM^{abcd} \mbox{ } \mbox{ becomes manifestly a metric } \mbox{ } \mM_{AB}
\ee 
and 
\be
\mN_{abcd} \mbox{ } \mbox{ becomes } \mbox{ } \mN^{AB} \mbox{ } . 
\ee
\ni Thus Dirac beables or observables for geometrodynamics require extra equation
\be
\mbox{\bf \{} ( \scH | \mJ ) \mbox{\bf ,} \, ( \iD_{\sfD} | \chi^{\sfD} ) \mbox{\bf  \}} \mbox{ } `='  \mbox{ } 0 \mbox{ } , 
\ee 
which gives 
\beq
\left(
\left\{
\frac{\delta \iD_{\sfD}}{\delta \underline{\mp}}
\left\{
\underline{\mG} - \underline{\underline{\mM}} \, \underline{\mD}^2  
\right\}
 -    \frac{\delta \iD_{\sfD}}{\delta \underline{\mh}} \, \underline{\underline{\mN}} \mbox{ } \underline{\mp} 
\right\}
\mL \mbox{\Huge |} \chi^{\sfD} 
\right)  
\label{42}
\eeq
for DeWitt vector quantities 
\beq
\underline{\mG} := \mbox{$\frac{2}{\sqrt{\sh}}$}   \big\{ \mp^{ia}  {\mp_a}^j  -  \mbox{$\frac{\sp}{2}$}\mp^{ij}  \big\} 
                 - \mbox{$\frac{1}{2\sqrt{\sh}}$}  \big\{ \mp^{ab}  \mp_{ab}   -  \mbox{$\frac{\sp^2}{2}$}        \big\} \mh^{ij}
                 - \mbox{$\frac{\sqrt{\sh}}{2}$}   \big\{ \mh_{ij}  \mR        -  2 \, \mR^{ij}  \big\}                                +   \sqrt{\mh} \, \Lambda \, \mh^{ij}
\eeq
familiar from ADM evolution \cite{ADM} and 
\be 
\underline{\mD}^2	:= \mD^i\mD^j \mbox{ } .
\ee 
In unsmeared form, (\ref{42}) is the FDE 
\beq
\{    \underline{\mG} -  \underline{\underline{\mM}} \mbox{ } \underline{\mD}^2    \}      \frac{\delta \iD_{\sfD}}{\delta \underline{\mp}} \mbox{ } `='  \mbox{ } 
                      2 \,  \underline{\mp}  \mbox{ }  \underline{\underline{\mN}}  \mbox{ }  \frac{\delta \iD_{\sfD}}{\delta \underline{\mh}}               \mbox{ } .
\eeq
Some examples of Dirac observables in GR for more specialized highly symmetric cases can be found in e.g. \cite{Marolf, Wada, TorreGowdy}.  

\mbox{ } 

\ni{\bf Acknowledgments} To those close to me gave me the spirit to do this.  
Thanks also to Chris Isham, Malcolm MacCallum, Jeremy Butterfield, Don Page, Enrique Alvarez and the Foundational Questions Institute.  

\begin{appendices}

\section{Supporting Lemmas}

\ni {\bf Lemma 1}. If $\iB_{\sfB}$ are beables, then so are the functionals ${\cal F}[\iB_{\sfB}]$.

\mbox{ }  

\ni \underline{Proof}.
For $L$ linear, if $u$ solves 
\be 
L \phi \mbox{ } `=' \mbox{ } 0 \mbox{ } , 
\ee 
then $f(u)$ and $F[u]$ also solve this equation by the chain-rule. 
Then indeed, the PDE or FDE for the beables is an equation of this form.  

\mbox{ }

\ni Remark 1) This renders `basis beables' a useful concept: i.e. a {\it sufficient set} of \K beables to describe one's theory by $K_{\sfK}$.   
These are a mutually functionally independent choice of 
\be 
\mbox{dim(reduced phase space)} = 2 \, q - c
\ee 
reduced phase space quantities (for $q = \mbox{dim}(\FrQ)$ and $c$ the amount of phase space degrees of freedom taken out by the constraints).
This is $2\{q - g\}$ in the simplest case of $g$ first-class constraints alone.
See e.g. \cite{FileR, QuadI} for RPM examples of `basis beables'.  

\mbox{ }

\ni Remark 2) (Composition Principle) In the case of multiple functional dependency restrictions applying, the composition of these restrictions applies; see Sec 2 for examples.

\mbox{ }

\ni {\bf Lemma 2}. The pure-configuration \K beables equation is the same as the equation determining which quantities are annihilated by the generators. 

\mbox{ }

\ni Remark 3) This refers to the configuration space $\FrQ$ uplift of the group generators acting upon absolute space $\FA$, 
by which one passes from $\FA$-invariants to ones built out of multiple particle positions $q^{iI}$.

\mbox{ } 

\ni \underline{Proof}.  This occurs since the constraints involved are homogeneous linear in $p_{iI}$, so the Poisson bracket removes the $p_{iI}$ factors  
and replaces them with $\pa/\pa q^{iI}$ factors. 
[The Poisson bracket term with the other sign is annihilated by the pure-configuration restriction.]

\mbox{ } 

\ni Remark 4) The pure-momentum case of \K beables has no such result as constraints are not confined to be linear in the $Q^A$. 
Those that are double-linear have pure-$P_{\sfA}$ and pure-$Q^{\sfA}$ beables close parallel.
On the other hand, those which are not have more divergence between the forms of these two contributions to the `basis beables'.

\end{appendices}


\end{document}